\documentclass[sigconf, nonacm]{acmart}

\usepackage{booktabs}
\usepackage{subcaption}
\usepackage{float}
\usepackage{xspace}
\newcommand{\stagezero}{Data Exploration\xspace}
\newcommand\vldbyear{2026}
\newcommand\vldbworkshop{DASHSys: Systems for Data-centric Agents with Human-in-the-loop}
\newcommand\vldbauthors{\authors}
\newcommand\vldbtitle{\shorttitle}

\newcommand\vldbpagestyle{plain}

\definecolor{lin}{HTML}{e31a1c}

\definecolor{yike}{HTML}{1f78b4}

\begin{document}

\title[Walk Before You Run]{Walk Before You Run: The Importance of \stagezero for Data Analysis Agents}

\author{Yike Yuan}
\affiliation{%
  \institution{University of Michigan}
  \city{Ann Arbor}
  \state{Michigan}
  \country{USA}
}
\email{yikeyuan@umich.edu}

\author{Virum Ranka}
\affiliation{%
  \institution{University of Michigan}
  \city{Ann Arbor}
  \state{Michigan}
  \country{USA}
}
\email{virum@umich.edu}

\author{Tina Lasisi}
\affiliation{%
  \institution{University of Michigan}
  \city{Ann Arbor}
  \state{Michigan}
  \country{USA}
}
\email{tlasisi@umich.edu}

\author{Lin Ma}
\affiliation{%
  \institution{University of Michigan}
  \city{Ann Arbor}
  \state{Michigan}
  \country{USA}
}
\email{linmacse@umich.edu}

\begin{abstract}
LLM-based data-analysis tools are increasingly used to help users analyze messy spreadsheets and workbooks, from answering questions over uploaded files to generating code, summaries, and visualizations. 
Existing tools include chat-with-data interfaces and code-executing analysis agents that inspect uploaded files, run analysis code, and return final outputs.
These systems are often evaluated by the correctness of their final downstream answers. 
However, reliable data analysis also depends on an earlier step: understanding what the dataset contains before solving the requested task. 
For complex workbooks, this \emph{\stagezero} step includes identifying the logical tables behind the physical sheets, interpreting column semantics, recovering keys and relationships, and detecting quality issues. 
In current tools and benchmarks, this step is usually left implicit: a system may produce a plausible final answer without exposing whether it has actually formed a correct, inspectable understanding of the dataset. 
This creates a gap between downstream task performance and the dataset understanding needed for reliable, human-checkable analysis.

Our key contribution is to identify this overlooked gap, make \stagezero a first-class evaluation target, and show through downstream experiments that stronger \stagezero support improves task performance. To evaluate dataset understanding directly, we introduce two benchmark settings: a real multi-sheet workbook benchmark based on a Vitamin D study dataset, and an extension of DSBench with schema-fixed \stagezero artifacts. In both settings, systems are evaluated by the quality of a structured artifact capturing tables, columns, semantic roles, relationships, and profiling signals. Our results show that strong LLMs and data-analysis agents still miss important logical structure even when they read spreadsheet content. Furthermore, explicit \stagezero support often improves downstream correctness — suggesting it should be treated as a first-class, inspectable stage in LLM data-analysis workflows and a natural human-in-the-loop checkpoint where domain experts can review and correct the artifact before downstream analysis proceeds.

\end{abstract}
\maketitle

\pagestyle{\vldbpagestyle}
\begingroup\small\noindent\raggedright\textbf{VLDB Workshop Reference Format:}\\
\vldbauthors. \vldbtitle. VLDB \vldbyear\ Workshop: \vldbworkshop.\\ 
\endgroup
\begingroup
\renewcommand\thefootnote{}\footnote{\noindent
This work is licensed under the Creative Commons BY-NC-ND 4.0 International License. Visit \url{https://creativecommons.org/licenses/by-nc-nd/4.0/} to view a copy of this license. For any use beyond those covered by this license, obtain permission by emailing \href{mailto:info@vldb.org}{info@vldb.org}. Copyright is held by the owner/author(s). Publication rights licensed to the VLDB Endowment. \\
\raggedright Proceedings of the VLDB Endowment. 
ISSN 2150-8097. \\
}\addtocounter{footnote}{-1}\endgroup

\section{Introduction}

LLM-based data-analysis tools are increasingly used to help users work with messy spreadsheets and workbooks: a user uploads a file, asks a natural-language question, and expects the system to return reliable answers, calculations, plots, tables, or summaries~\cite{jing2024dsbench}.
Reliable analysis in this setting is challenging because many real tasks begin not from a clean relational database, but from spreadsheet files whose logical structure is only partially explicit.
In multi-sheet workbooks, the visible layout of cells may not coincide with the logical data objects needed for analysis: one sheet may contain several tables, a wide matrix may encode a normalized relation, repeated period columns may imply a long fact table, and categorical fields may serve as reusable dimensions.

Existing LLM-based data-analysis tools already make this workflow much easier, including chat-with-data interfaces, spreadsheet agents, and autonomous data-science agents~\cite{pandasai_github,li2023sheetcopilot,openai_ada_doc,zhang2025deepanalyze}.
They can inspect uploaded files, generate or execute analysis code, and return downstream outputs from natural-language requests.
However, these workflows usually expose the final analysis result, not a separate representation of what the system believes the dataset contains before analysis begins.
This leaves a key part of the analysis process implicit: whether the system has formed a faithful understanding of the dataset before solving the user's task.

We call this prerequisite phase \textbf{\stagezero}: the process of constructing a dataset-grounded understanding before downstream analysis begins.
For complex workbooks, \stagezero includes identifying logical tables behind physical sheets, interpreting column semantics, recovering keys and relationships, and detecting lightweight quality signals.
If this pre-analysis understanding is wrong, later reasoning can still look plausible while being grounded in the wrong schema.
Conversely, even when a system answers correctly, an end-to-end score alone may not reveal whether it understood the dataset faithfully or arrived at the answer through brittle reasoning.
This makes failures hard to localize and successes hard to trust.

\begin{figure}[tph]
    \centering
    \includegraphics[width=\linewidth]{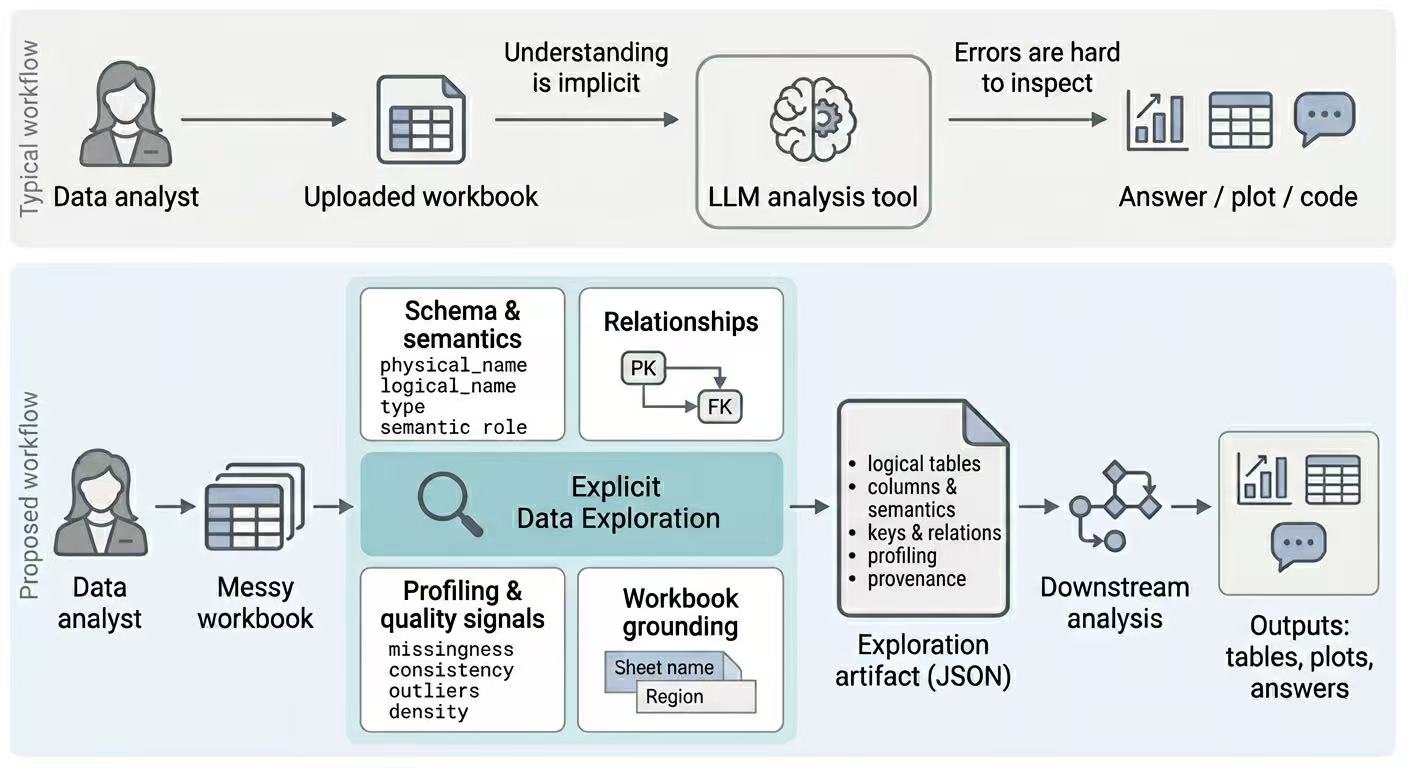}
    \caption{Typical LLM data-analysis workflows versus our proposed workflow with explicit \stagezero.}
    \label{fig:diagram}
\end{figure}

Figure~\ref{fig:diagram} illustrates the distinction.
The common workflow moves directly from an uploaded workbook and user question to downstream execution and final outputs.
Our proposed workflow inserts an explicit \stagezero artifact before downstream analysis: a structured, dataset-grounded representation of the system's understanding of the workbook.
The goal is not merely to make the model explain its reasoning, but to externalize what it has inferred about the dataset in a form that can be inspected, corrected, reused, and evaluated independently.
This also creates a concrete human-in-the-loop checkpoint: a domain expert who understands what the data represents — but may not write analysis code — can review and correct the artifact before downstream reasoning commits to a potentially flawed dataset interpretation.

A similar gap appears in evaluation.
Recent benchmarks have made substantial progress in measuring LLM data-analysis capability.
For example, DS-1000 studies data-science code generation with automatic evaluation~\cite{lai2022ds1000}, InfiAgent-DABench evaluates question-driven data analysis over CSVs~\cite{hu2024infiagent}, DataSciBench studies end-to-end data-science workflows~\cite{zhang2025datascibench}, and DSBench pushes realism with competition-derived data-analysis tasks~\cite{jing2024dsbench}.
These benchmarks are valuable, and current systems can score impressively on them: the ChatGPT agent, for instance, reports 89.9\% pass@1 on DSBench's data-analysis tasks, surpassing the human baseline of 64.1\%~\cite{openai2025chatgptagent}.
Yet these evaluations are still primarily centered on downstream task success.
They usually ask whether the system produced the right answer, code, or artifact, rather than whether it first built a correct and inspectable understanding of the dataset.
As a result, the \stagezero gap remains under-tested.

This paper studies that gap by making \stagezero an explicit object of evaluation.
We develop two benchmark settings for evaluating dataset understanding directly.
First, we construct a multi-sheet workbook benchmark based on a real-world Vitamin D study dataset acquired from the anthropology department in our institution, where reliable analysis requires understanding repeated measurements, participant identifiers, timepoints, non-data rows, and measurement-quality structure.
Second, we extend 12 DSBench data-analysis tasks with schema-fixed \stagezero artifacts and ground-truth annotations for messy-workbook datasets, while keeping the original downstream questions unchanged.
In both settings, systems are evaluated by the quality of a structured artifact that records logical tables, columns, semantic roles, relationships, profiling signals, and source grounding.

We use these benchmarks and downstream experiments to test whether the \stagezero gap matters in practice.
Our benchmark results show that the hardest failures are not surface reading errors, but mistakes in logical schema recovery, implicit entity identification, and relation inference.
In downstream tasks, we vary the amount and reliability of \stagezero support available before task execution.
Across the real workbook and selected DSBench tasks, stronger \stagezero support often improves downstream correctness, although the gains depend on whether the system actually uses and re-grounds the artifact.

The main lesson is that \stagezero should not be treated as an unobserved byproduct of downstream reasoning.
It is a distinct source of both capability and failure: a system can fail because it misunderstands the dataset before analysis begins, and it can improve when that understanding is made explicit.
This also has implications for human-in-the-loop data-analysis agents.
A structured \stagezero artifact gives domain experts a concrete object to inspect, correct, and reuse before downstream analysis proceeds.
Rather than positioning the human only as a query writer or final-output reviewer, explicit \stagezero creates an earlier intervention point where human judgment about variables, entities, relationships, and data quality can directly shape the analysis.

\paragraph{Contributions.}
This paper makes the following contributions:
\begin{itemize}
  \item \textbf{Benchmark settings for explicit \stagezero evaluation.}
  We develop a real multi-sheet workbook benchmark based on a Vitamin D study dataset and extend 12 DSBench data-analysis tasks with schema-fixed \stagezero artifacts, ground-truth annotations, prompts, and an automatic evaluation pipeline for the main structural, semantic, relational, and grounding components of \stagezero.

  \item \textbf{Evidence that current systems have persistent \stagezero gaps.}
  Evaluating representative LLMs and data-analysis agents shows that dataset understanding remains difficult even when systems can read workbook contents.
  The largest gaps arise in logical schema recovery, implicit entity identification, and relation inference, rather than in surface-level extraction alone.

  \item \textbf{Downstream evidence that \stagezero support improves analysis.}
  Through control/middle/treatment experiments on the real workbook and 12 DSBench tasks, we show that explicit or oracle Data Exploration support often improves downstream correctness and helps localize failures — and because the artifact is inspectable before execution begins, it also gives domain experts a natural point to intervene before misunderstandings propagate.
\end{itemize}
To support reproducibility, we release the code, prompts, schema template, evaluation pipeline, and benchmark assets at \url{https://github.com/coconut0621/walk-before-you-run}.

\section{Background and Motivation}
\label{sec:background}
\paragraph{\normalfont\bfseries Data analysis over messy workbooks.}
We focus on LLM-based data analysis in which a system must turn uploaded spreadsheet files and a natural-language request into reliable downstream outputs, such as answers, tables, plots, cleaned datasets, feature files, or summaries.
Unlike question answering over a single clean table, realistic workbook tasks require the system to infer the relevant data objects, valid records, field meanings, and cross-sheet relationships before performing the requested analysis.
Examples include answering questions over a financial model, cleaning triplicate measurements from a study workbook, or computing statistics from a multi-sheet spreadsheet whose logical tables are not explicit.
These tasks are common in real analysis workflows, and mistakes in this early interpretation step can silently affect later calculations.
\paragraph{\normalfont\bfseries LLM development for data analysis: tool use and reasoning.}
Recent progress in LLM-based data analysis has been driven in part by two capabilities: tool use and inference-time reasoning.
With tool use, models can invoke external functions, inspect uploaded files, execute analysis code, and retrieve information during task solving~\cite{schick2023toolformer, yao2023react, openai2025chatgptagent}.
With inference-time reasoning, models can decompose a request into intermediate steps before producing a final answer~\cite{openai2024o1systemcard, deepseek2025r1}.
Both capabilities are valuable for downstream execution, but neither by itself guarantees dataset-specific structural understanding.
A model can write and run code over the wrong logical table, join fields that do not represent the intended entities, or treat a layout artifact as valid data.
Similarly, a fluent reasoning trace is not the same as an externally checkable account of the dataset: recent work shows that chain-of-thought rationales can diverge from a model's actual computational process~\cite{barez2025cot}.
For data analysis, these capabilities therefore need to be complemented by an explicit representation of dataset understanding.

\paragraph{\normalfont\bfseries Specialized LLM-based tools for data analysis.}
A growing ecosystem of LLM-based tools targets data analysis specifically. Systems such as PandasAI~\cite{pandasai_github}, SheetCopilot~\cite{li2023sheetcopilot}, and ChatGPT’s data-analysis workflow~\cite{openai_ada_doc} let users load or upload data and interact with it through natural-language requests. In these systems, dataset understanding is typically folded into the process of answering the user’s question and executing analysis code, rather than treated as a distinct stage of analysis.
DeepAnalyze~\cite{zhang2025deepanalyze}, one of the most recent purpose-built autonomous data science agents, makes planning more explicit before acting on the environment. However, that planning is still directed toward solving the downstream task, rather than producing a separable description of dataset structure that stands on its own as an object of inspection, correction, or scoring.
Across this landscape, a consistent pattern emerges: \emph{prior work rarely makes dataset understanding an explicit, standalone, benchmarked object.}
This motivates evaluating dataset understanding separately from downstream task execution.

\paragraph{\normalfont\bfseries What real data analysts do first.} 
Human data analysts typically treat dataset comprehension as a necessary early part of analysis~\cite{kandel2012enterprise, kandel2011wrangling}.
Wongsuphasawat et al.~\cite{wongsuphasawat2019eda} interviewed 18 professional data analysts and found that \emph{profiling}---assessing data quality, identifying structural properties, and building a mental model of the dataset---is the one exploration goal shared across \emph{all} analyses.
Unlike discovery (finding new insights), which mainly characterizes open-ended analyses, profiling appears across analytical settings, including those with a known downstream question.
Data science guides reflect the same view: exploratory analysis is iterative~\cite{wickham2023r4ds}. Even when the main question is already given, analysts still need to investigate what the data contains and whether its quality matches expectations~\cite{wongsuphasawat2019eda}.

In real analytical work, a dataset's logical structure is often not identical to its physical layout, especially in spreadsheets and multi-sheet workbooks.
The visible organization of cells may only partially reflect the analytical objects that matter downstream, so determining what the data actually contains requires inference beyond surface reading.

We therefore formalize this neglected step as \stagezero.

\section{\stagezero}
\label{sec:stage0}

\textbf{\stagezero} is the pre-analysis phase in which a system turns the physical layout of uploaded workbooks into an explicit logical description that downstream analysis can rely on.
In our setting, this description summarizes the workbook as analysis-ready metadata: the logical data objects behind sheets or blocks, their columns and semantic roles, primary and foreign keys, relationships, and lightweight quality/profiling signals.
Section~\ref{sec:benchmark-task} formalizes these categories as the schema-fixed artifact used in our benchmark.
It is not equivalent to reading sheets or extracting cell values.
A workbook may be readable at the cell level while still being logically misunderstood: the relevant analytical objects may be distributed across sheets, embedded in repeated blocks, or implicit in layout conventions rather than explicitly named as database tables.

For example, in our real Vitamin D study workbook, reliable downstream analysis requires more than locating measurement cells.
The system must recognize participant identifiers, summer and winter timepoints, triplicate skin-reflectance measurements, non-data rows, and measurement-quality structure before it can produce cleaned features.
Similarly, in messy DSBench workbooks, the analytical objects needed for a downstream question may be latent in spreadsheet layout rather than explicitly named as database tables.

We operationalize \stagezero as a \emph{schema-fixed structured artifact}: a single JSON object with a fixed schema that explicitly describes the dataset before downstream reasoning begins.
This artifact makes dataset understanding (i) \emph{comparable} across systems, (ii) \emph{auditable} and \emph{correctable} by domain experts before downstream execution begins, and (iii) \emph{automatically scorable} without requiring access to hidden internal states.
It also makes the ``walk before you run'' principle concrete: before solving downstream tasks, the system should already have identified the dataset objects, keys, relations, and profiling that downstream analysis will depend on.

\section{\stagezero Benchmark}
\label{sec:benchmark}
\begin{figure}[tph]
    \centering
    \includegraphics[width=1\linewidth]{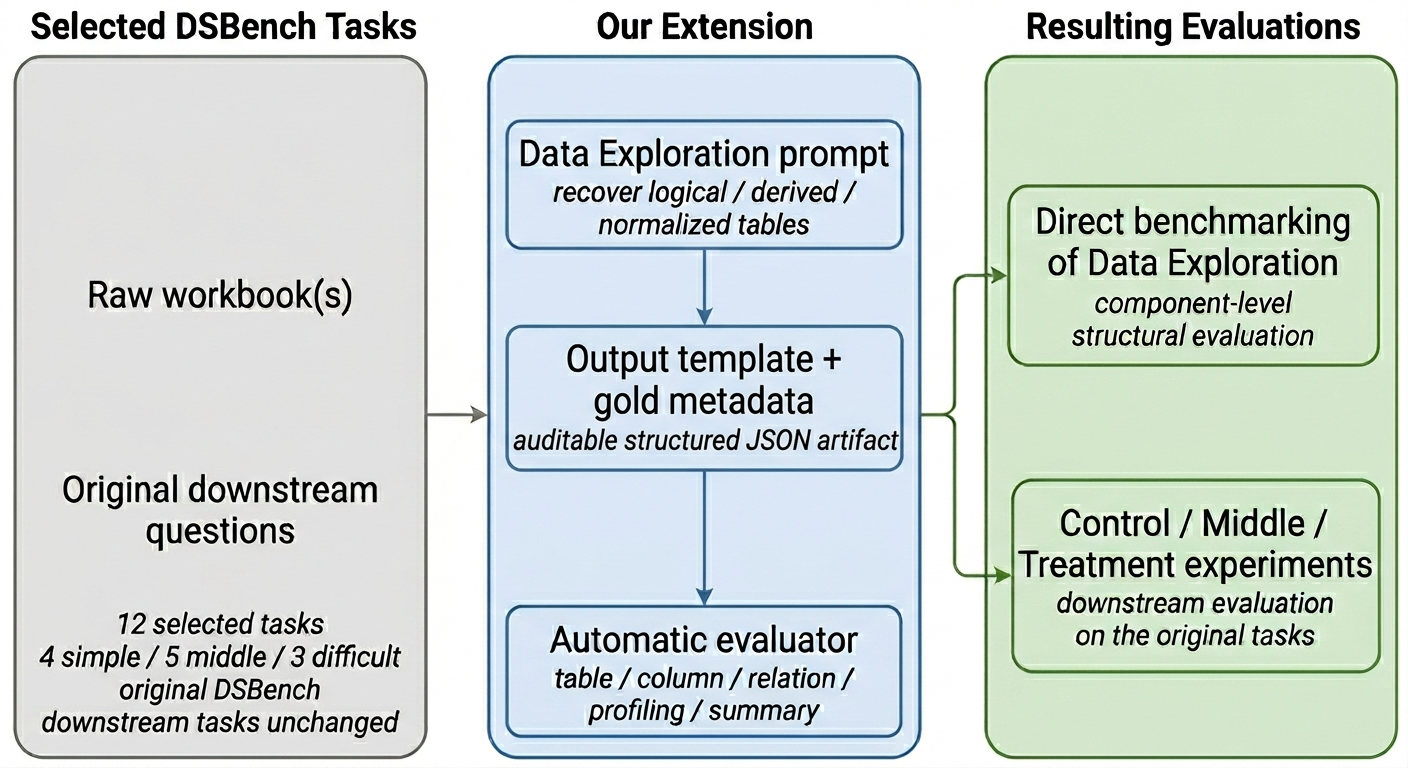}
    \caption{We build on raw workbooks and the original downstream questions, and add an explicit Data Exploration layer. }
    \label{fig:diagram2}
\end{figure}

Building on datasets from DSBench and a real-world Vitamin D study workbook acquired from the anthropology department in our institution, we directly measure dataset understanding by asking systems to produce a schema-fixed \stagezero artifact.
For the selected DSBench tasks, Figure~\ref{fig:diagram2} summarizes the extension: we keep the raw workbooks and original downstream questions unchanged, but add a \stagezero prompt, an output template, ground-truth metadata, and an automatic evaluator.

\subsection{Specification-driven \stagezero task}
\label{sec:benchmark-task}
Formalizing the schema-fixed structured artifact introduced in Section~\ref{sec:stage0}, the benchmark task asks the system to produce a conforming JSON output given one or more raw Excel workbooks and a prompt, \emph{before} any downstream analysis begins.
The target of inference is the dataset's \emph{logical} structure rather than the superficial workbook layout; the system is explicitly encouraged to recover normalized or derived tables when that better reflects the underlying data organization.
The required output includes:

\begin{itemize}
  \item \textbf{Dataset structure:} logical tables, their names, and row counts;
  \item \textbf{Schema and semantics:} columns with both \texttt{physical\_name} (the workbook-grounded field label) and \texttt{logical\_name} (the normalized canonical name), together with data types, semantic roles, and dictionary fields;
  \item \textbf{Relational structure:} primary keys, foreign keys, and inter-table relations over those logical tables;
  \item \textbf{Workbook grounding:} explicit table provenance through \texttt{source\_ref}, which records where the evidence came from in the workbook (e.g., sheet, nearby anchor text, source block, and approximate cell region);
  \item \textbf{Lightweight quantitative characterization:} table/column profiling fields, including density, parse success, type consistency, numeric summaries, date-format consistency when applicable, and robust outlier statistics for numeric columns;
  \item \textbf{Documentation:} a dataset-level structured summary.
\end{itemize}

This design makes \stagezero a schema-fixed inference task over \emph{messy workbook blocks}: systems are free to infer normalized logical structure, but must report it in an auditable format.

\subsection{Gold metadata format}

For each dataset, we maintain a compact human-editable ground-truth JSON in the same overall format as the required output.
The gold artifact records the expected logical tables, columns, relations, provenance fields, and profiling-relevant structure needed for evaluation.
The gold schema is written at the level of \emph{logical dataset entities}, not spreadsheet tabs: a wide matrix, repeated-period block, or parameter section may correspond to one or more derived relational tables.


\subsection{Multi-step evaluation pipeline}

After JSON validation, predicted and gold objects are aligned as equivalent logical objects may differ in naming or normalization. The table score $s_T$ estimates whether two tables represent the same logical object from structural, semantic, and provenance cues:
\[
s_T=0.42s_{\mathrm{col}}+0.10s_{\mathrm{row}}
+0.08(s_{\mathrm{sample}}+s_{\mathrm{role}})
+\alpha s_{T,\mathrm{name}}+\beta s_{\mathrm{src}},
\]
where $(\alpha,\beta)=(0.08,0.24)$ with provenance, prioritizing grounding, and $(0.32,0)$ without it, falling back on names. Tables are maximum-weight one-to-one matched and retained at $s_T\geq0.58$.

Within matched tables, column identity is scored by
\[
s_C=0.70s_{C,\mathrm{name}}+0.10\mathbb{I}_{\mathrm{dtype}}
+0.10\mathbb{I}_{\mathrm{role}}+0.05\mathbb{I}_{\mathrm{nullable}}
+0.05\mathbb{I}_{\mathrm{unique}},
\]
weighting names as the primary identity signal and the remaining properties as supporting cues. Pairs with $s_C\geq0.55$ are retained. 

Both scores are used only for alignment, and retained pairs receive unit credit.
For $x\in\{T,C,R\}$, $P_x$, $R_x$, and $F1_x$ denote precision, recall, and F1:
\[
P_x=|\mathcal{M}_x|/|\mathcal{P}_x|,\quad
R_x=|\mathcal{M}_x|/|\mathcal{G}_x|,\quad
F1_x=2P_xR_x/(P_x+R_x),
\]
where $\mathcal{G}_x$, $\mathcal{P}_x$, and $\mathcal{M}_x$ are the gold, predicted, and matched sets; relations are projected through alignment. Unmatched objects count as false positives or negatives. Columns are scored within matched tables and relations are direction-normalized. Exact fields use canonicalized matching; profiling uses tolerance-based numeric scores.

The raw score is the weighted mean of table/column/relation F1, data-type accuracy, and role accuracy (weight $1.0$ each); table/column logical-name accuracy ($0.5$ each); row-count accuracy ($0.4$); source similarity/all-fields accuracy ($0.7/0.3$); and profiling/outlier quality ($0.3/0.2$), omitting inapplicable terms.
\[
S=S_{\mathrm{raw}}\bigl[1-0.3(1-q)\bigr],
\]
where $q\in[0,1]$ is the mean LLM-judged summary coverage and faithfulness, capping the summary penalty at $30\%$.

\section{Benchmark Testing Results}
\label{sec:benchresults}

\subsection{Experimental setup}
\label{sec:bench-setup}
We evaluate \stagezero on one real multi-sheet Vitamin D workbook and a targeted subset of 12 DSBench data-analysis tasks, selected because their workbooks contain nontrivial multi-sheet and
spreadsheet-specific structures (e.g., multiple logical blocks, matrix/repeated layouts, and cross-sheet dependencies). The subset targets this setting rather than representing the full
DSBench.

We group them into 4 simple tasks (05, 19, 29, and 38), 5 middle tasks (01, 18, 22, 25, and 27), and 3 difficult tasks (08, 10, and 43), based on the original downstream requirements and workbook reasoning involved: \emph{simple} tasks are mostly multiple-choice questions over relatively direct workbook structure; \emph{middle} tasks require combining messier multi-sheet structures or intermediate workbook logic; and difficult tasks require recovering implicit multi-entity or matrix structures (08 and 10), or progressively repairing linking, dragging, and circular-reference errors (43).
We use the same split for both the direct \stagezero benchmark and the downstream experiments.
The Vitamin D workbook provides a real-world setting with repeated measurements, multiple timepoints, non-data rows, and domain-specific measurement structure.

All systems are given the raw workbook files, the same \stagezero prompt, and the same schema-fixed output template.
They are asked to produce a structured JSON artifact before any downstream task execution.
This section evaluates the quality of that artifact directly; the control/middle/treatment downstream experiments are introduced separately in Section~\ref{sec:experiments}.

We evaluate Gemini~3.1~Pro~\cite{gemini31pro}, Claude~Opus~4.6~\cite{claudeopus46}, and GPT~5.4 \cite{gpt54} on the selected benchmark settings. We enable the highest thinking/adaptive capabilities when possible.
We also evaluate GPT's agent mode~\cite{openai2025chatgptagent} where applicable.
Because agent mode has usage limitations and is less suitable for large batch evaluation, we run it only on the difficult DSBench group and on the Vitamin D workbook.
All non-agent systems are run in the same direct \stagezero generation mode: they receive the workbook input and produce the required JSON artifact without access to gold metadata or downstream answers.

Quality is measured with the automatic evaluator described in Section~\ref{sec:benchmark}. The reported component scores cover table, column, relation, profiling, grounding, and summary quality, together with an overall structural score.
For DSBench, we report component scores averaged within each difficulty group.
For the Vitamin D workbook, we report component scores on the single real-world workbook instance.

\subsection{Results on selected DSBench tasks}

\begin{figure}[tbp]
\centering
\begin{subfigure}{\linewidth}
  \centering
  \includegraphics[width=0.92\linewidth]{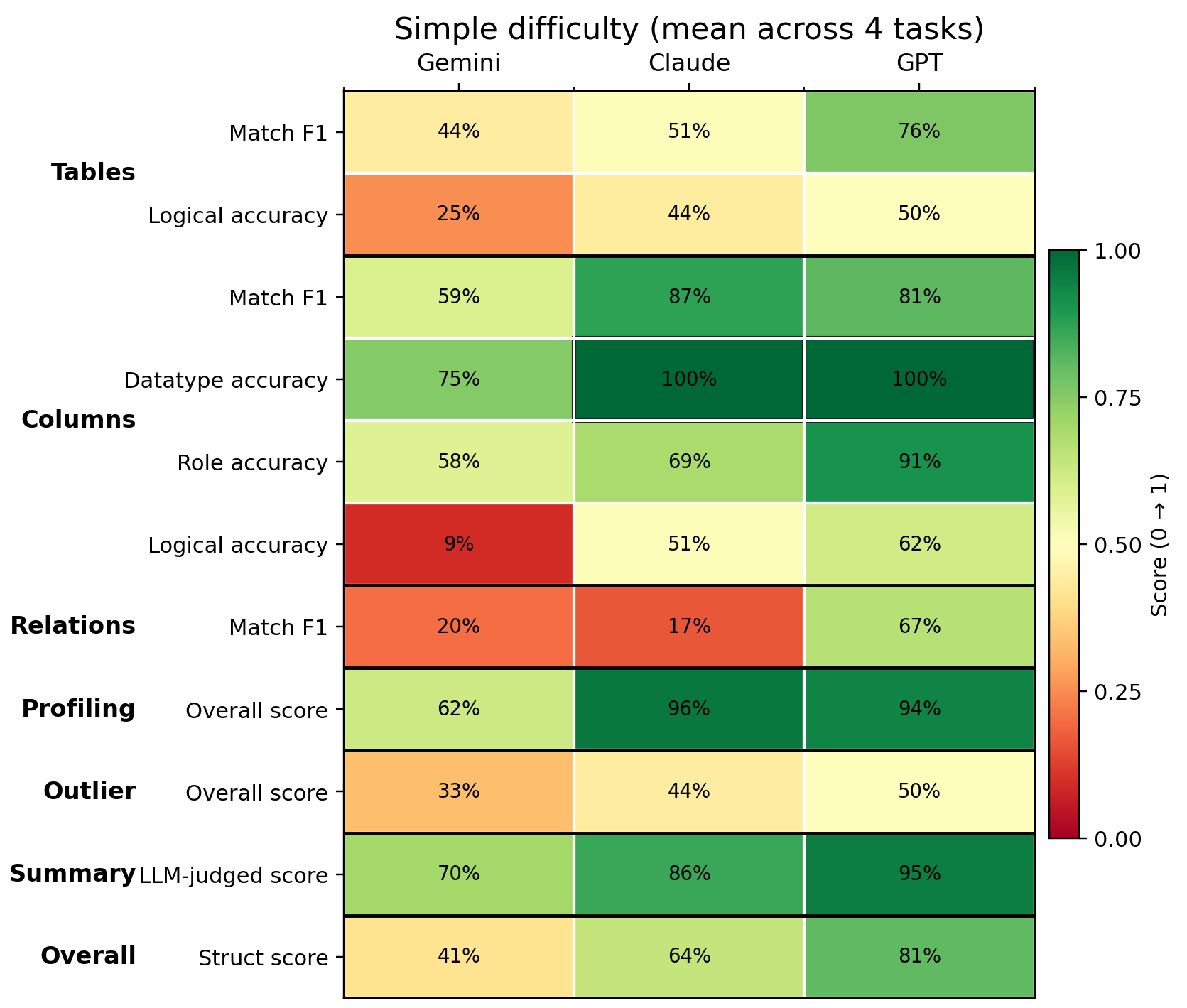}
\end{subfigure}

\vspace{0.4em}

\begin{subfigure}{\linewidth}
  \centering
  \includegraphics[width=0.92\linewidth]{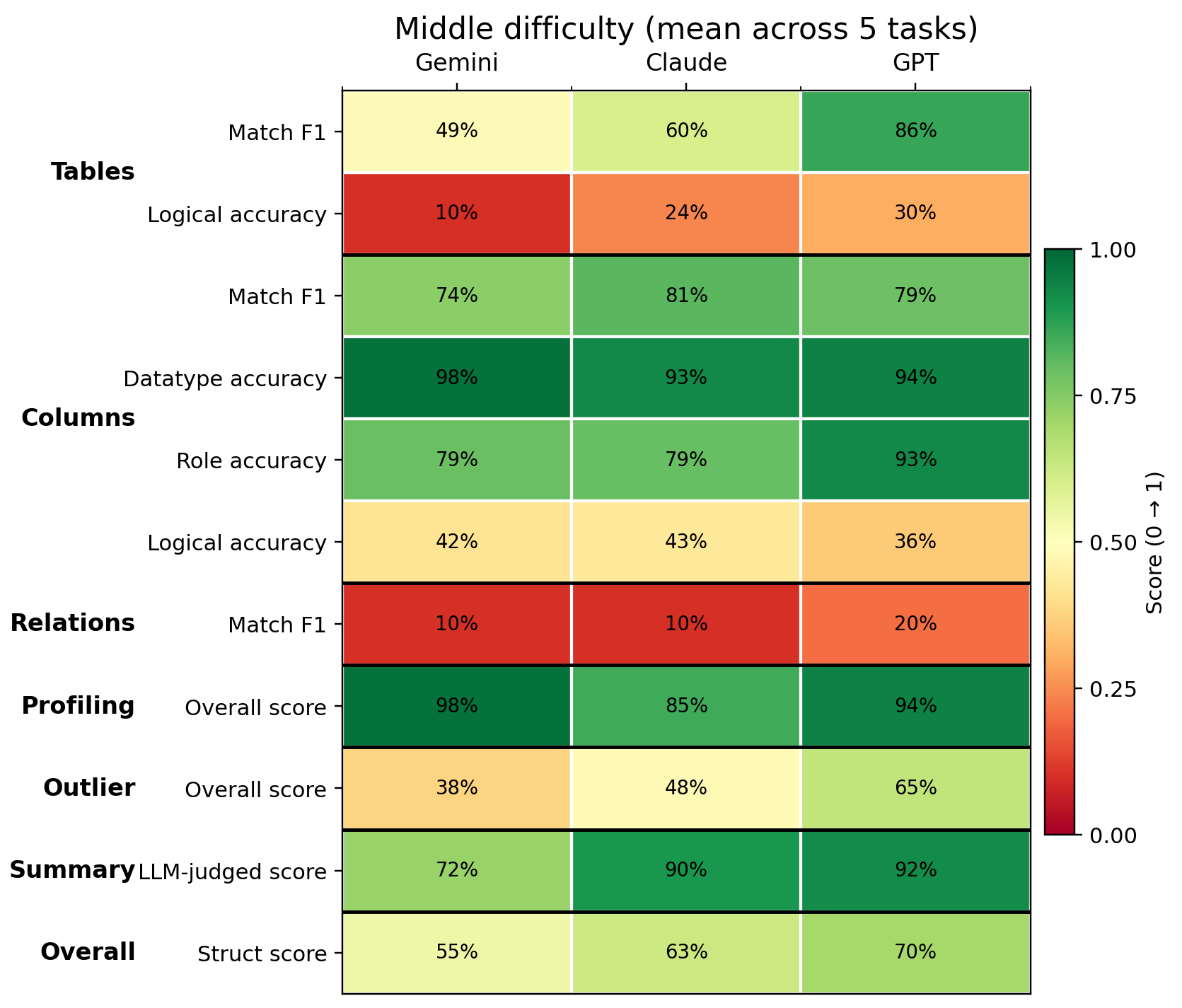}
\end{subfigure}

\vspace{0.4em}

\begin{subfigure}{\linewidth}
  \centering
  \includegraphics[width=0.92\linewidth]{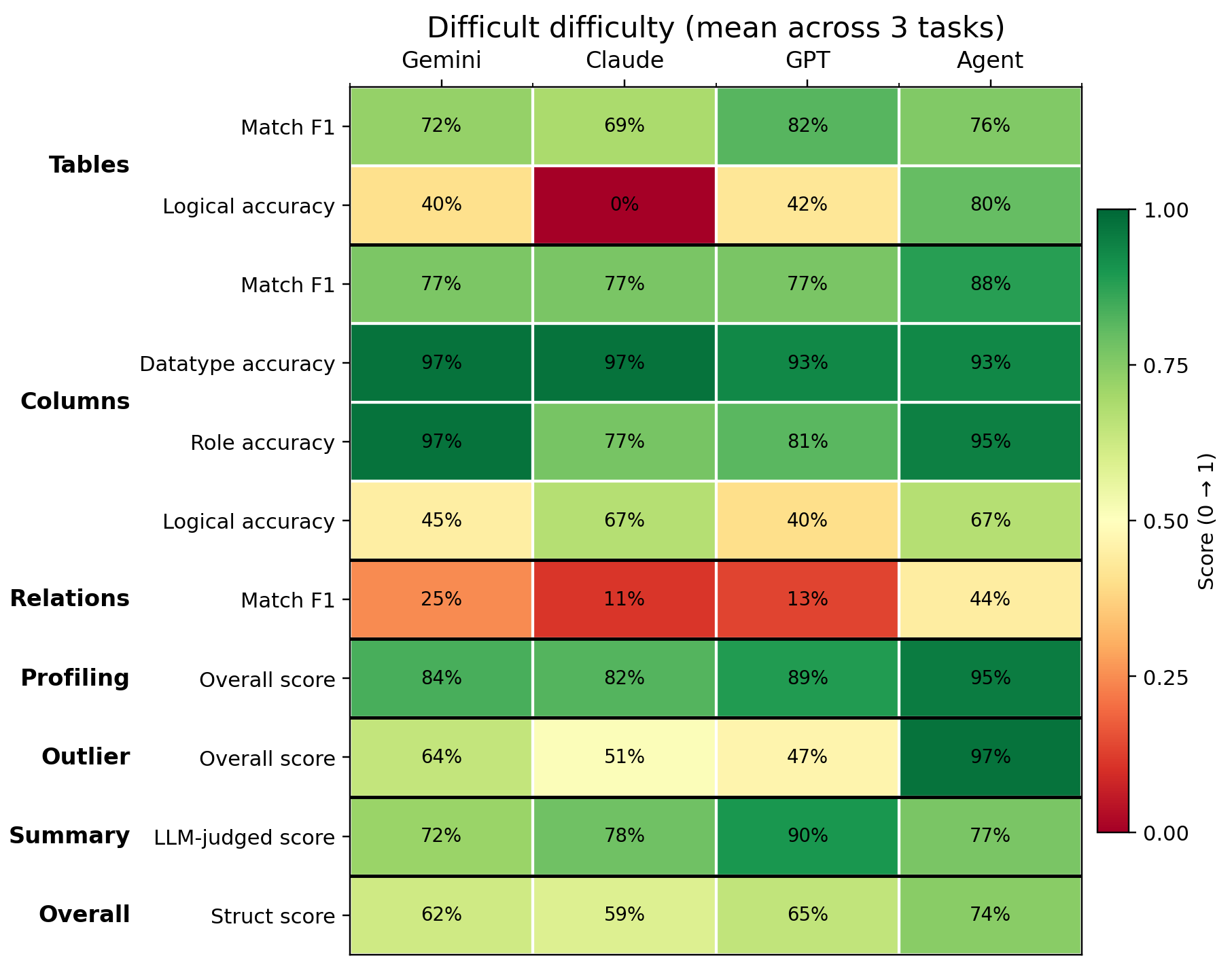}
\end{subfigure}
\caption{Component-level \stagezero benchmark scores on selected DSBench datasets, including table/column/relation matching and logical accuracy, profiling, outlier handling, LLM-judged summary quality, and overall structural score. Each panel reports means within one difficulty group.}
\label{fig:dsbench-stage0-heatmap}
\end{figure}

Figure~\ref{fig:dsbench-stage0-heatmap} reports component-level \stagezero scores on the 12 selected DSBench tasks, averaged within the simple, middle, and difficult groups.
The difficult panel additionally includes GPT's agent mode on the same three difficult tasks.

\paragraph{Key findings.}
First, \stagezero remains nontrivial even on benchmark tasks that are ultimately designed for downstream analysis: across systems, the largest and most persistent gaps appear in \emph{logical} understanding and \emph{relation} recovery, rather than in surface-level table or column extraction.
Second, system quality is clearly component-dependent. Some systems are relatively strong on structural recovery or summaries, while still leaving substantial errors in latent schema recovery, especially when workbook regions encode multiple entities or implicit objects.
Third, as task difficulty increases, \stagezero weaknesses become more varied and more structurally meaningful, motivating the following case studies, which show how these failures arise in realistic workbook layouts.

\paragraph{Case study: missed decomposition in a multi-entity workbook block.}
This workbook is the dataset for task 08 in DSBench, where the block \emph{Vehicles for service and hub groupings} encodes several logical entities in one visible spreadsheet region. Because the prompt asks systems to recover \emph{logical / derived / normalized} tables rather than simply transcribe sheets, the gold \stagezero decomposes this region into \texttt{Scenarios}, \texttt{Hubs}, and \texttt{ScenarioDepotInputs}. This separates reusable scenario definitions and hub entities from the scenario--depot facts that drive downstream allocation.

Claude partially reconstructs this structure. It correctly identifies a normalized scenario--depot fact table and also recovers several explicit blocks, including depot capacity, depot distance, and parameter tables. However, it collapses the scenario--hub region into one table, \texttt{depot\_scenario\_vehicles}, with fields such as \texttt{scenario\_id}, \texttt{depot}, \texttt{vehicles\_for\_service}, and \texttt{hub\_allocation}. This is not just a naming or formatting difference. In the gold schema, \texttt{Scenarios} and \texttt{Hubs} are separate logical objects that anchor the intended foreign-key structure, whereas Claude's prediction treats hub membership as attributes inside a single fact-like table.

This mistake explains the relation-level failure: Claude predicts only one relation, while the gold \stagezero contains five. The model reads many visible workbook blocks correctly, but misses part of the latent relational decomposition needed to represent the workbook as an analytical dataset.

\paragraph{Case study: missing an implicit table hidden behind a matrix block.}
Task 10 in DSBench uses a football-season workbook with a match-results table, auxiliary team-level sheets, and a bonus matrix over team pairs. The key challenge is that one analytically important object is not presented as a conventional row-wise table. Instead, the \texttt{Bonus} sheet encodes pair-level information through a matrix-style layout, where the logical rows should correspond to unordered team pairs and their series-level outcomes.

GPT recovers several row-wise structures. It identifies \texttt{Matches} from the \texttt{Data} sheet, \texttt{Teams} from \texttt{Check\_Sum}, and the per-team stop-time table from \texttt{Q16}, showing that it can read ordinary table-like workbook regions. However, it does not recover the \texttt{Bonus} sheet as a separate pair-level table. Instead, it predicts \texttt{Team\_Primes} while missing the more consequential table linking two teams with bonus and winner information.

This omission affects both table and relation scoring. In the gold \stagezero, the pair-level table should link each bonus record to the two participating teams and to the winner field. Because GPT misses the table, these relations are absent as well. The error is therefore not a local column mistake or a superficial naming mismatch; it is a failure to infer an implicit logical object from a non-row-wise workbook layout. This case complements the previous one: even when visible tables are recovered accurately, \stagezero can still fail if the system does not reconstruct latent objects encoded by spreadsheet-specific structure.

\subsection{Real-world Vitamin D workbook}

\begin{figure}[tbp]
    \centering
    \includegraphics[width=1\linewidth]{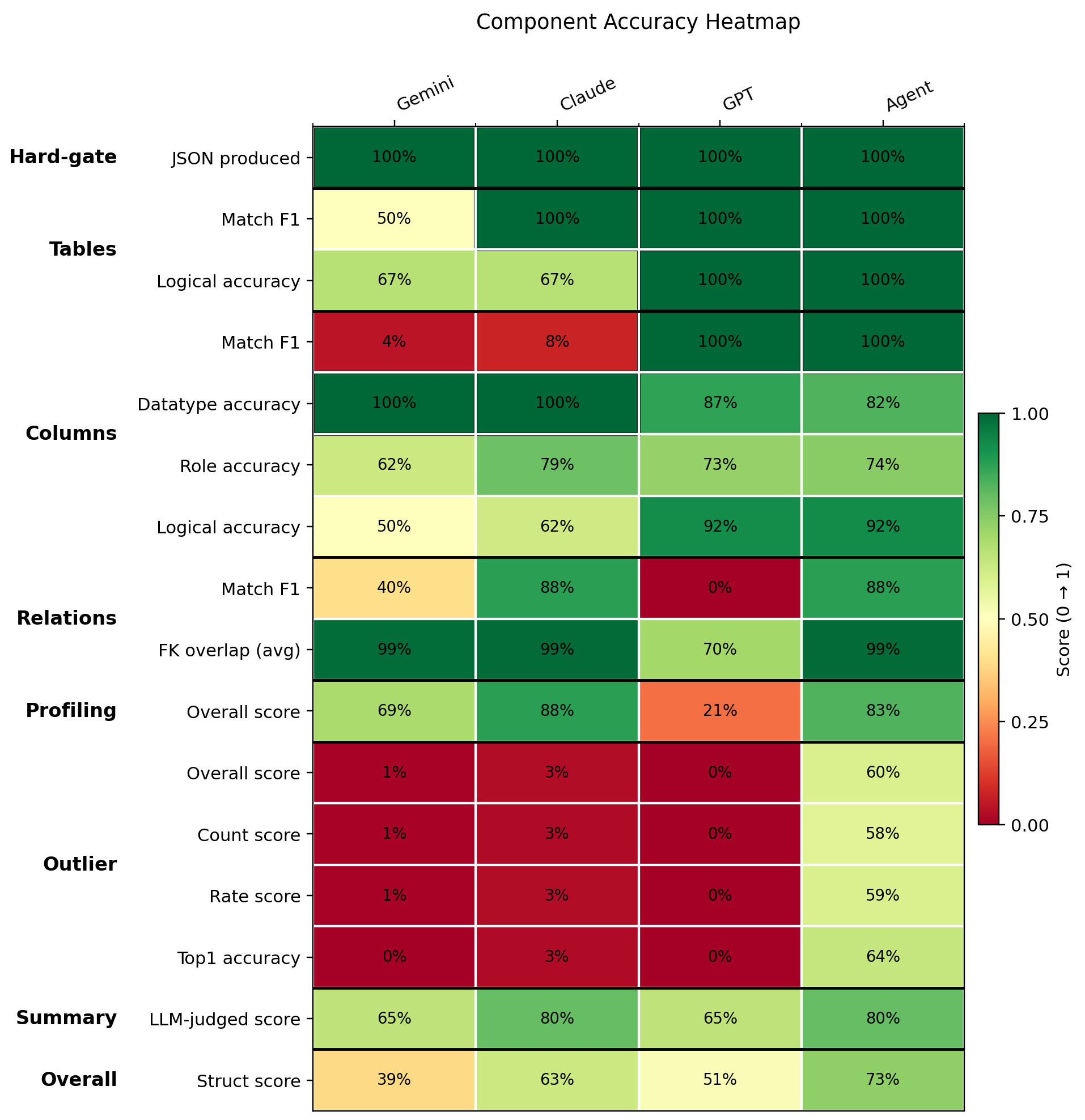}
    \caption{Component-level \stagezero benchmark scores on the real-world workbook. Scores include table/column/relation matching and logical accuracy, profiling, outlier handling, LLM-judged summary quality, and overall structural score.}
    \label{fig:vitamind-stage0-heatmap}
\end{figure}

Figure~\ref{fig:vitamind-stage0-heatmap} reports component-level \stagezero scores for the real Vitamin D workbook.
Low relation-recovery and lightweight quantitative-characterization scores show that \stagezero difficulty is not unique to the DSBench extension: consequential failures persist in real analytical workbooks. We next test whether stronger \stagezero support is associated with better downstream performance.

\section{Does \stagezero Improve Downstream Analysis?}
\label{sec:experiments}

\subsection{Experimental setup}

We evaluate whether explicit \stagezero support improves downstream data-analysis performance on the same two families of workbook settings used in Section~\ref{sec:benchresults}: the real Vitamin D study workbook and the 12 selected DSBench data-analysis tasks.
Unlike Section~\ref{sec:benchresults}, which scores the \stagezero artifact itself, this section evaluates final downstream task correctness.

For each dataset and downstream task, we compare three experimental conditions summarized in Table~\ref{tab:conditions}.
All conditions share the same downstream questions and scoring; they differ only in the \stagezero support available before answering.
The oracle artifact used in \textsc{Treatment} contains only pre-analysis metadata, not downstream answers, task-specific derivations, or solution hints.
The goal is not to hold token usage constant across conditions, but to test whether inserting an explicit \stagezero step or artifact improves later analysis.

\begin{table}[tph]
\centering
\small
\begin{tabular}{p{0.15\linewidth} p{0.77\linewidth}}
\toprule
\textbf{Condition} & \textbf{What the system gets before downstream questions} \\
\midrule
\textsc{Control} & Raw files only \\
\textsc{Middle} & Raw files + the system's own \stagezero JSON (generated from the files) \\
\textsc{Treatment} & Raw files + oracle \stagezero JSON (ground truth) \\
\bottomrule
\end{tabular}
\caption{\stagezero as an intervention: three conditions used in our downstream experiments.}
\label{tab:conditions}
\end{table}

For the Vitamin D workbook, we solicit real-world analysis questions from its provider.
The downstream task requires systems to consolidate triplicate skin-reflectance measurements into cleaned participant--timepoint feature tables for \textit{summer} and \textit{winter}.
We report accuracy on objective dataset-understanding questions and generated feature-table accuracy against the expert reference.

For DSBench, we keep the original downstream questions unchanged and report answer accuracy under the three conditions, using the same 4/5/3 difficulty split defined in Section~\ref{sec:bench-setup}.
The \stagezero artifact is auxiliary metadata for downstream reasoning, not a disguised answer key.

We evaluate the same main model versions as in the direct \stagezero benchmark: Gemini~3.1~Pro~\cite{gemini31pro}, Claude~Opus~4.6~\cite{claudeopus46}, and GPT~5.4~\cite{gpt54}.
We also evaluate GPT's agent mode~\cite{openai2025chatgptagent} on the difficult DSBench group and on the Vitamin D workbook.
For the simple DSBench group, we additionally include DeepAnalyze and ai-analyst~\cite{zhang2025deepanalyze,ai_analyst_github}; due to tool limitations, these systems are not tested on the middle or difficult groups.

\subsection{Experimental results}

\begin{figure}[tbp]
    \centering
    \includegraphics[width=0.82\linewidth]{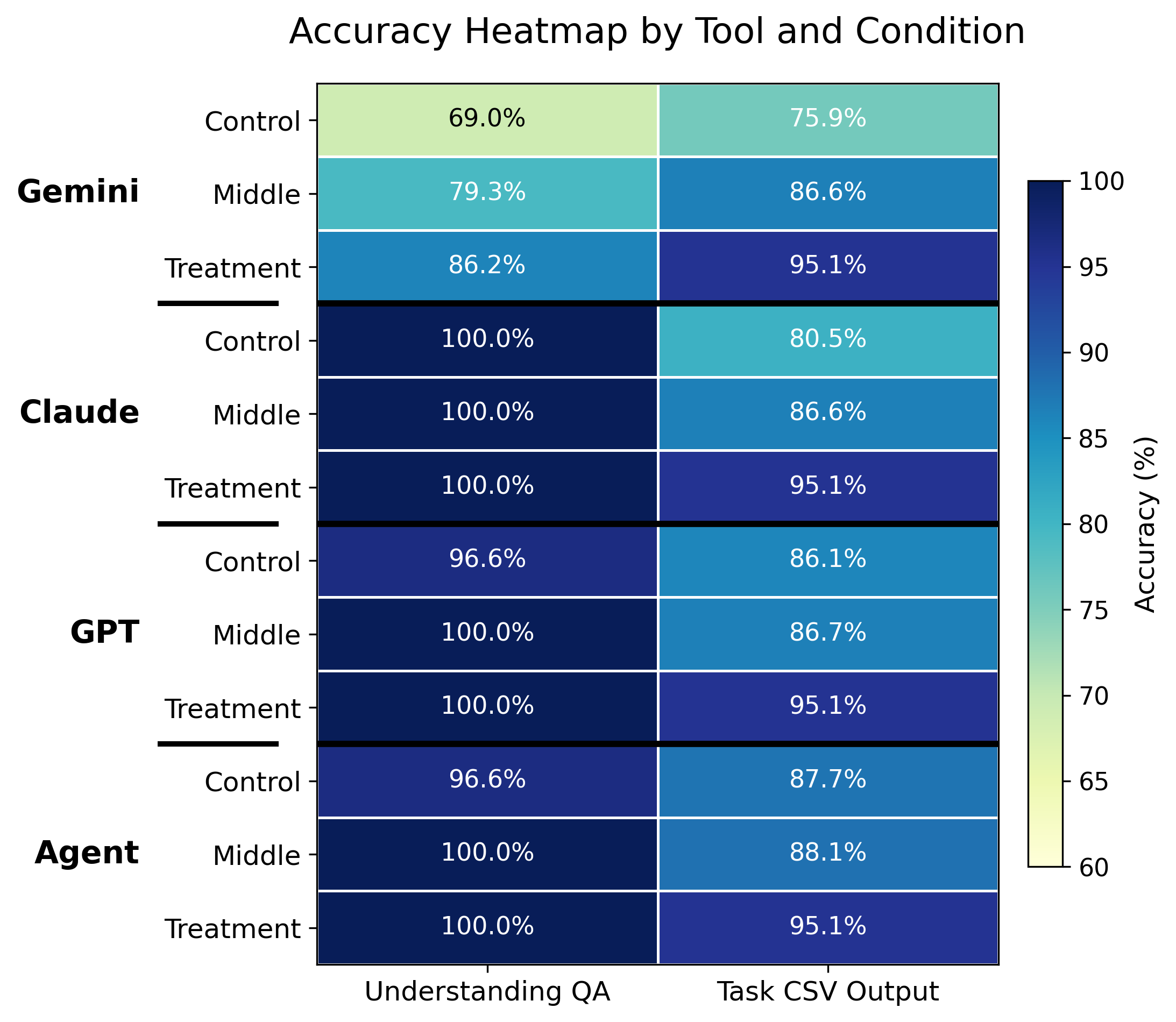}
    \caption{Accuracy on the real Vitamin D workbook experiments. Rows show tools and prompting conditions; the left column reports accuracy on dataset-understanding questions, and the right column reports accuracy on the downstream feature-table task (outputs evaluated as CSV files).}
    \label{fig:real-workflow-heatmap}
\end{figure}

\begin{figure}[tbp]
    \centering
    \includegraphics[width=1\linewidth]{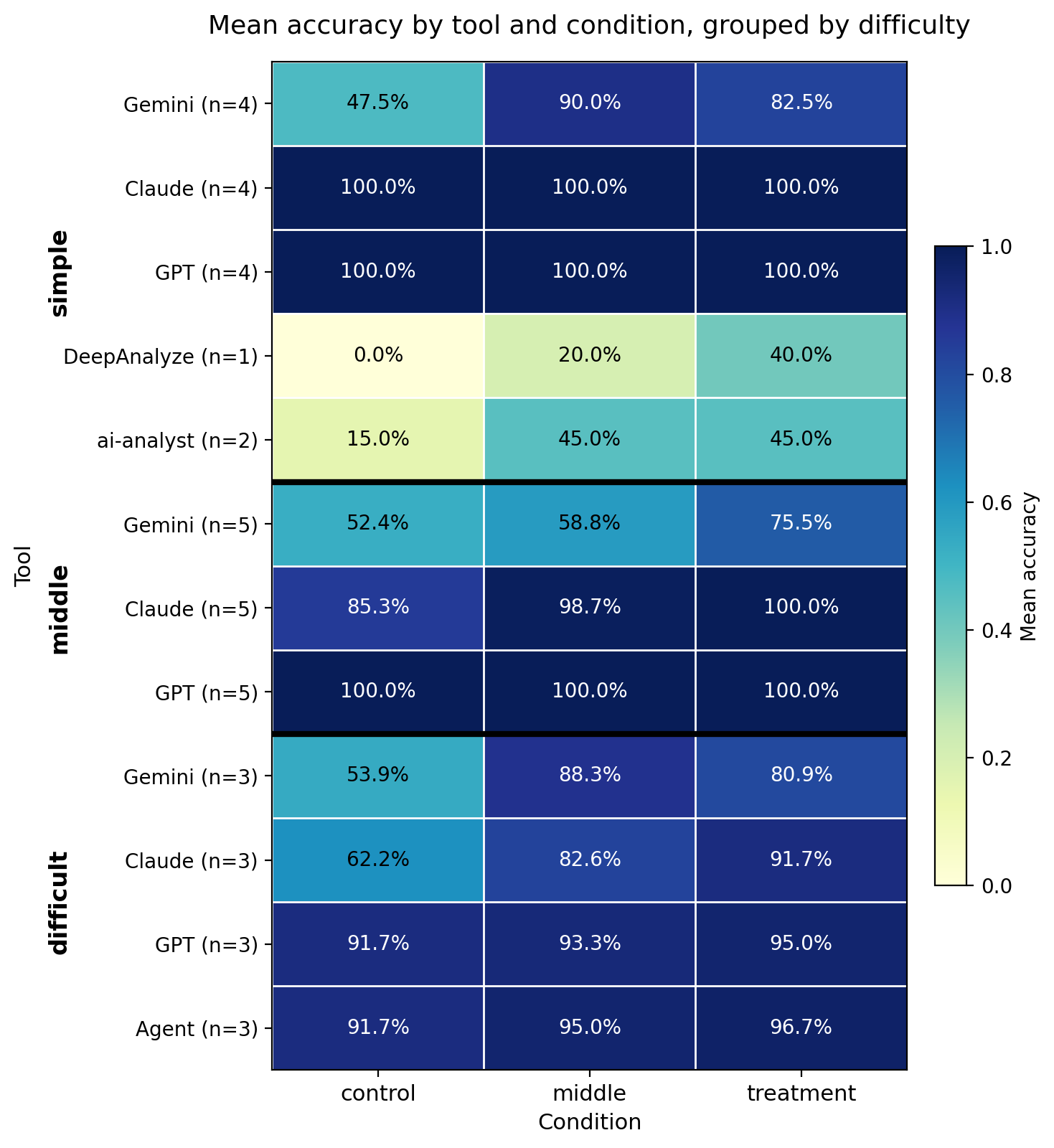}
    \caption{Mean accuracy of different tools in different conditions, grouped by difficulty.}
    \label{fig:dsbench-grouped-heatmap}
\end{figure}

Figures~\ref{fig:real-workflow-heatmap} and~\ref{fig:dsbench-grouped-heatmap} summarize the downstream results on the real workbook workflow and the selected DSBench tasks.
Across both settings, stronger \stagezero support is generally associated with better downstream correctness.

\paragraph{Real-world workbook task.}
On the real Vitamin D workbook, performance increases from \textsc{control} to \textsc{middle} and then to \textsc{treatment}, especially on the downstream feature-table output.
The gains are most pronounced for Gemini, while others are already strong on the understanding questions and still benefit from stronger \stagezero support on the downstream task.
Even when a system can answer many schema-level questions correctly, explicit \stagezero support still helps translate that understanding into more reliable execution on the real analysis workflow.

\paragraph{Selected DSBench tasks.}
On DSBench, we observe the same qualitative pattern but with substantially larger variation across task difficulty.
\stagezero support is most helpful on the middle and difficult groups, where success depends less on surface extraction and more on recovering latent structure, relations, and intermediate spreadsheet logic before answering the final questions.
Although \textsc{treatment} is usually the strongest condition, we still observe occasional cases where \textsc{middle} slightly outperforms \textsc{treatment}, suggesting that externally provided \stagezero information is not always fully leveraged unless the model first constructs the representation itself.

\paragraph{Case study: a difficult financial-model task.}
This task is built on a multi-sheet financial workbook with deliberate linking, dragging, and circular-reference errors. The instructions explicitly require the solver to progressively fix each error identified before answering the
final downstream questions, so later answers depend on reconstructing cross-sheet dependencies rather than checking isolated formulas. GPT improves from \textbf{15/20} in \textsc{Control} to \textbf{16/20} in \textsc{Middle} and \textbf{17/20} in \textsc{Treatment}.

The clearest \textsc{Control}$\rightarrow$\textsc{Middle} gain appears on Q18 and Q20, the final downstream questions. In \textsc{Control}, GPT treats them as uncertain end-of-chain questions and selects incorrect options. In \textsc{Middle}, after first reconstructing the workbook's dependency structure, GPT recognizes that the debt schedule depends on beginning cash, operating and investing cash flow, and debt repayment logic, and that debt issuance/repayment should not itself enter ``cash available for debt repayment'' because that would create circularity. This changes Q18 to b and Q20 to a.

\textsc{Treatment} adds a more localized benefit on Q1, where GPT uses the oracle artifact to identify Debt row 6---the ``Cash flow available for debt repayment'' line, which links to row 26---as the source of circularity. GPT still misses Q3 and Q6, which shows that \stagezero support improves dependency reasoning and error localization, but does not eliminate all formula-level errors.

The same task also separates \stagezero availability from \stagezero use. On Q8, GPT thinking mode answers b in both \textsc{Middle} and \textsc{Treatment}, defaulting to the generic accounting rule that EBIT should explicitly deduct D\&A. GPT agent mode instead answers the gold option a, recognizing that in this workbook depreciation and amortization are already embedded in operating costs. Thus the remaining challenge is not only recovering workbook structure, but using it for workbook-specific semantic judgments.

\paragraph{Case study: a depot-allocation task.}
This task asks the model to simulate MVC's depot-allocation procedure across scenarios. It requires combining several workbook structures: scenarios alter hub partitions, hubs contain different depots, depots have service capacities, and overflow vehicles are greedily reassigned to the shortest feasible trip. Correct answers thus require reconstructing the latent procedure encoded across these blocks.

Claude improves monotonically, from \textbf{6/9} in \textsc{Control} to \textbf{7/9} in \textsc{Middle} and \textbf{9/9} in \textsc{Treatment}. The main \textsc{Control}$\rightarrow$\textsc{Middle} gain is on Q30, which asks for Hub~1's total transport cost in Scenario~1b. In \textsc{Control}, Claude recognizes that Hub~1 expands to depots A--D but does not complete the reassignment chain and selects \texttt{H}. In \textsc{Middle}, after reconstructing the scenario and hub structure, it follows the correct overflow pattern: \emph{D}$\rightarrow$\emph{A}, \emph{C}$\rightarrow$\emph{A}, and \emph{B}$\rightarrow$\emph{A}; 56 vehicles are penalized, yielding the correct option \texttt{I}. The gain is therefore not a simple arithmetic correction, but comes from recovering how scenario membership, depot capacity, distance ordering, and greedy reassignment interact.

The \textsc{Middle}$\rightarrow$\textsc{Treatment} gain is different. On Q27 and Q33, Claude in \textsc{Middle} computes the correct quantities but maps them to the wrong multiple-choice letters. With the oracle \stagezero artifact, it retains the computations but grounds the final selections correctly. This case thus separates two benefits of stronger \stagezero support: \textsc{Middle} improves procedural decomposition, while \textsc{Treatment} reduces residual answer-grounding errors.

The same task provides a counterexample: Gemini rises from \textbf{6/9} in \textsc{Control} to \textbf{9/9} in \textsc{Middle}, then falls to \textbf{7/9} in \textsc{Treatment}. The reversal is on Q31 and Q32 for Hub~2 in Scenario~1b. In \textsc{Middle}, Gemini correctly applies \emph{I}$\rightarrow$\emph{H}, \emph{E}$\rightarrow$\emph{F}, then \emph{I}$\rightarrow$\emph{F} with \emph{F}'s remaining capacity, leaving 20 vehicles unserviced and yielding the correct Q31 revenue of \$5{,}202{,}000. In \textsc{Treatment}, it instead uses \emph{I}$\rightarrow$\emph{H}, \emph{E}$\rightarrow$\emph{G}, and \emph{I}$\rightarrow$\emph{F}, producing the error and carrying it into Q32.

Receiving a \stagezero artifact does not ensure that every downstream step is re-grounded in the workbook. Producing it in \textsc{Middle} may force closer inspection of the table and reassignment rule, whereas \textsc{Treatment} can fail under incomplete procedural verification. The effect is not uniformly negative---\textsc{Treatment} still repairs Q29 by correcting a local transport-cost discrepancy---but shows that \stagezero support must be actively used.

\section{Discussion and Future Work}
Our evidence suggests \stagezero is both practically important and currently under-evaluated, motivating it as an additional benchmark axis alongside downstream question answering.
Future work includes expanding dataset diversity, more directly quantifying how specific \stagezero components relate to downstream gains and failure modes, and exploring how \stagezero artifacts can be cached, edited, and reused as persistent structured memories. Although an explicit \stagezero step adds upfront cost, that cost can be amortized: once a dataset-specific artifact has been produced or verified, multiple downstream tasks can reuse the same structured context, potentially reducing repeated schema re-inference, token cost, and hallucinated joins.

\paragraph{Broader implications: from pipelines to human-in-the-loop analysis.}
Beyond benchmarking, \stagezero surfaces a deeper point about the nature of data analysis itself.
Most data analysis tasks do not have a single correct answer or a single correct solution path---the space of possible questions to ask, and the interpretations to draw, depends critically on what the data actually contains.
If we frame AI-assisted analysis as a pipeline that begins from a pre-specified question, we leave little room for human agency: the human poses a query, the system executes it.
But if a \stagezero artifact is produced and made available before final downstream execution, the interaction changes fundamentally.
Humans---including domain experts who understand what the data represents but may lack programming fluency---can engage at the stage where their knowledge matters most: deciding what to look for, identifying which variables are meaningful, flagging quality issues that a schema-level summary makes visible.
Data analysis then shifts from a question-answering pipeline toward a collaborative process in which human creativity and domain expertise are not bypassed but actively engaged.
In this sense, \stagezero is not only a prerequisite for accurate downstream computation; it is also an auditable entry point through which domain experts can re-enter the analytical loop.

\section{Conclusion}
Reliable LLM data-analysis tools should ``walk before they run'': before answering downstream questions, they should form a faithful, dataset-grounded understanding of the workbook.
This paper makes that pre-analysis step explicit as \stagezero.
We develop benchmark settings for evaluating \stagezero directly, including a real multi-sheet Vitamin D workbook and an extension of selected DSBench tasks with schema-fixed \stagezero artifacts and automatic component-level evaluation.
Our results show that current LLMs and data-analysis agents still have measurable \stagezero gaps, particularly in recovering implicit logical structure and relationships from messy workbooks.
We further show through downstream experiments that stronger \stagezero support often improves task correctness, while also revealing that models must actively use and re-ground the artifact for the benefit to appear.
By making \stagezero explicit, inspectable, and evaluable, we aim to move AI-assisted data analysis from a closed query-to-answer pipeline toward a workflow in which both model reliability and human oversight can improve.

\begin{acks}
This work was supported (in part) by the Google ML and Systems Junior Faculty Award, the Google JAX AI Stack Research Award, and the NVIDIA Academic Grant Program Award.
\end{acks}

\newpage

\bibliographystyle{ACM-Reference-Format}
\bibliography{refs}

\end{document}